\documentclass[showpacs,twocolumn,amsmath,superscriptaddress]{revtex4-1}
\usepackage[dvipdfmx]{graphicx}
\usepackage{color,amsmath,lmodern,bm,array,braket}

\newcommand{\cH}{\mathcal{H}}

\newcolumntype{C}[1]{>{\centering\arraybackslash}p{#1}}
\newcolumntype{L}[1]{>{\raggedright\arraybackslash}p{#1}}
\newcolumntype{R}[1]{>{\raggedleft\arraybackslash}p{#1}}

\begin{document}

\preprint{Lichtenstein}

\title{Formation mechanism of helical $\bm Q$ structure in Gd-based skyrmion materials
%: A first-principles study
}%

\author{Takuya Nomoto}%
\email{nomoto@ap.t.u-tokyo.ac.jp}%
\affiliation{Department of Applied Physics, The University of Tokyo, Hongo, Bunkyo-ku, Tokyo, 113-8656, Japan}%
\author{Takashi Koretsune}%
\affiliation{Department of Physics, Tohoku University, Miyagi 980-8578, Japan}%
\author{Ryotaro Arita}%
\affiliation{Department of Applied Physics, The University of Tokyo, Hongo, Bunkyo-ku, Tokyo, 113-8656, Japan}
\affiliation{RIKEN Center for Emergent Matter Science (CEMS), Wako 351-0198, Japan}
\date{\today}%

%Abstract
\begin{abstract}
Using the {\it ab initio} local force method, we investigate the formation mechanism of the helical spin structure in GdRu$_2$Si$_2$ and Gd$_2$PdSi$_3$. We calculate the paramagnetic spin susceptibility and find that the Fermi surface nesting is not the origin of the incommensurate modulation, in contrast to the naive scenario based on the Ruderman-Kittel-Kasuya-Yosida mechanism. We then decompose the exchange interactions between the Gd spins into each orbital component, and show that spin-density-wave type interaction between the Gd-5$d$ orbitals is ferromagnetic, but the interaction between the Gd-4$f$ orbitals is antiferromagnetic. We conclude that the competition of these two interactions, namely, the inter-orbital frustration, stabilizes the finite-$\bm Q$ structure. 
%of the skyrmion crystal formations. 
\end{abstract}

\maketitle

{\it Introduction.}
A magnetic skyrmion, topologically-protected swirling spin texture, has recently attracted much attention because of its potential application to a spintronics memory. Among the features suitable for an information bit, the size of each skyrmion in the skyrmion crystal 
is essential for designing high-density storage devices.

In a non-centrosymmetric crystal, it is widely believed that competition between the ferromagnetic exchange interaction $\mathcal{J}$ and Dzyaloshinskii-Moriya (DM) interaction $\mathcal{D}$ stabilizes a magnetic skyrmion~\cite{Nagaosa2013}. Since the DM interaction prefers a spatially twisted spin configuration, its helical period $\lambda$ is determined by the ratio $\mathcal{J}/\mathcal{D}$, which is around 20-100~nm in typical skyrmion crystals, including MnSi~\cite{Ishikawa1976,Muhlbauer2009}, FeGe~\cite{Lebech1989,Yu2011}, and Cu$_2$OSeO$_3$~\cite{Adams2012,Seki2012}. The latest first-principles estimations of $\mathcal{J}/\mathcal{D}$ show good agreement with $\lambda$ in the experiments for these materials~\cite{Gayles2015, Koretsune2015, Kikuchi2016}. 

On the other hand, more recently, formation of a skyrmion crystal has been reported for several Gd-based centrosymmetric compounds. Although the atomic configuration of Gd atoms in these compounds has a diverse variety such as the frustrated triangular lattice in Gd$_2$PdSi$_3$~\cite{Kurumaji2019}, breathing Kagome lattice in Gd$_3$Ru$_4$Al$_{12}$~\cite{Max2019}, and body-centered tetragonal lattice in GdRu$_2$Si$_2$~\cite{Seki2020}, they have a common helical period around 2-3 nm, which is much shorter than those originated from the DM interaction. 

From the theoretical point of view, there are several possible mechanisms for stabilizing a skyrmion crystal in centrosymmetric materials~\cite{Okubo2012, Leonov2015, Lin2016, Batista2016, Heinze2011, Ozawa2017, Hayami2017, Hayami2014, Wang2019}. For example, the role of the geometrical frustration of the short-range two-spin interactions~\cite{Okubo2012, Leonov2015, Lin2016, Batista2016} and the four-spin interactions mediated by the itinerant electrons~\cite{Heinze2011, Ozawa2017, Hayami2017} have been investigated. While the former theory predicts the skyrmion formation in frustrated systems, the latter predicts it even in the high-symmetric non-frustrated crystals. The short period skyrmions observed in the non-centrosymmetric crystals MnGe~\cite{Kanazawa2011}, EuPtSi~\cite{Kakihana2018, Kaneko2019}, and Y$_3$Co$_8$Sn$_4$~\cite{Takagi2018} have also been discussed in terms of the mechanisms which do not rely only on the DM interaction~\cite{Okumura2019, Brinker2019, Grytsiuk2020}.

Despite many successes of these theories for the short period skyrmions, the actual mechanism in these materials are still under debate. This is mainly due to the lack of microscopic calculation based on the realistic electronic structure.
%to validate their scenarios. 
Indeed, most of the known skyrmion crystals with short periods are metallic magnets, and thus, the inherent exchange interactions should be sensitive to the details of the electronic property. Since the period of the skyrmion crystal is essentially the same as that of the helical spin phase which often appears as the ground state of the skyrmion materials, it is important to understand what stabilizes the short period modulation, characterized by the modulation vector $\bm Q$, in the helical state.

In this paper, we study the helical spin structure in GdRu$_2$Si$_2$ and Gd$_2$PdSi$_3$ from first-principles.
%focusing on the modulation vector $\bm Q$. 
We use the local force approach to obtain the exchange interactions $J_{ij}$ and discuss the stable structures within mean-field level. For GdRu$_2$Si$_2$, not only the calculated modulation vector $\bm Q$ but also the N\'eel temperature $T_N$ show good agreement with the experiments. Then, we decompose $J_{ij}$ into each orbital component 
%and perform spin susceptibility calculations for the paramagnetic state 
to reveal the mechanism to stabilize the finite-${\bm Q}$ modulation. Our analysis shows that while the spin-density-wave (SDW) type interaction between the Gd-5$d$ electrons is ferromagnetic, the interaction between the Gd-4$f$ electrons is antiferromagnetic.
The finite-$\bm Q$ structure originates from the competition between these interactions, namely, the inter-orbital frustration inherent in the Gd spins. 
On the other hand, through the calculation of the paramagnetic spin susceptibility, we find that the Fermi surface nesting is not the origin of the incommensurate modulation. We also show that the conventional Ruderman-Kittel-Kasuya-Yosida (RKKY) interaction becomes negligibly small and does not play any role in this compound. 
We observe a similar behavior also in Gd$_2$PdSi$_3$, indicating that the frustration between the $d$ and $f$ channel is ubiquitous, at least in the Gd-based skyrmion materials. 

{\it Methods.}
We start with calculations for GdRu$_2$Si$_2$ and Gd$_2$PdSi$_3$
based on the generalized gradient approximation (GGA) within spin density functional theory (SDFT). We use WIEN2k code~\cite{wien2k}, where we assume the collinear ferromagnetic order~\cite{note01,supple}. Here, we employ the exchange correlation functional proposed by Perdew, Burke, and Ernzerhof~\cite{pbe}, and neglect the spin-orbit coupling since the orbital moment of the Gd-4$f$ orbital is quenched. Then, we obtain the tight-binding model in the Wannier representation by using {\sc Wannier90} code~\cite{w90,w903} through {\sc wien2wannier} interface~\cite{w2w}. Finally, the exchange interaction $J_{ij}$ is evaluated by the following formula in the local force approach~\cite{Oguchi1983, Lichtenstein1984, Korotin2015, Nomoto2020}:
\begin{align}
J_{ij}=-\frac{1}{2}{\rm Tr}_{\omega_n\ell\sigma}[G_{ji}\Sigma^{i0}G_{ij}\Sigma^{j0}],\label{eq:lich}
\end{align}
where $G_{ij}$ and $\Sigma^{i0}$ are the Green's function and magnetic potential perturbation due to the spin rotation, respectively. In this paper, we employ the local approximation for $\Sigma^{i0}$~\cite{Nomoto2020, note04} and use the intermediate representation for the Matsubara frequency summation~\cite{Shinaoka2017, Chikano2019, Li2019}. The Fourier transform
$J({\bm q})$ of Eq.~\eqref{eq:lich} gives the most stable spin structure in the mean-field level, and the corresponding $T_N({\bm q})$ is estimated by $T_N({\bm q})=2J({\bm q})/3$. The details of the calculation are given in Ref.~\cite{supple}. 

\begin{figure}[t]
\centering
\includegraphics[width=8.5cm]{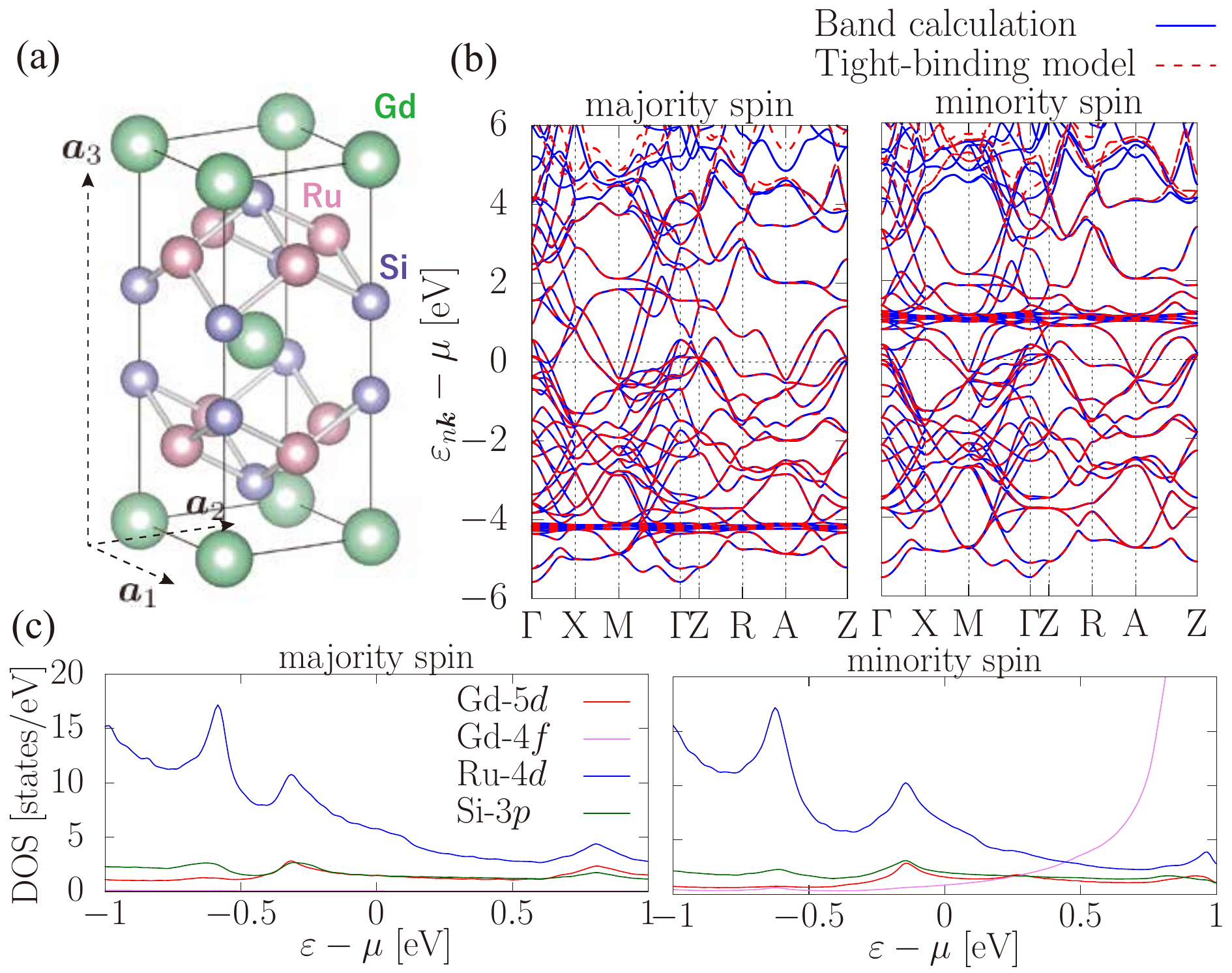}
\caption{(a) The crystal structure, (b) band structure, and (c) partial density of states of GdRu$_2$Si$_2$. In (b), the results obtained based on SDFT calculation are indicated by the solid blue lines and those on the Wannier tight-binding model by the red dashed lines. }
\label{fig:1} 
\end{figure}
\begin{figure}[t]
\centering
\includegraphics[width=8.0cm]{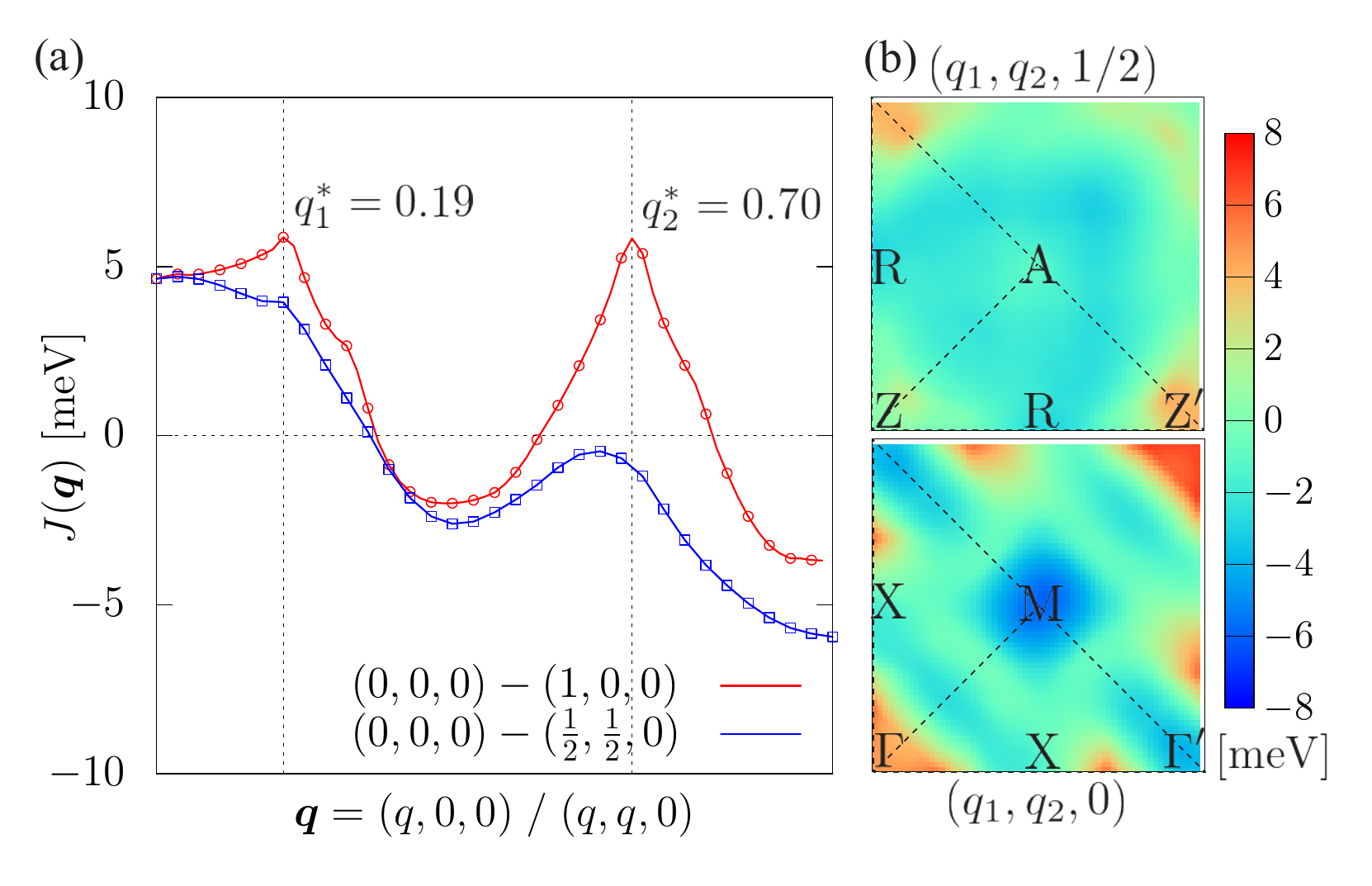}
\caption{The momentum dependence of $J({\bm q})$ (a) along the symmetry lines (0,0,0)-(1,0,0) (red) and (0,0,0)-($\frac{1}{2},\frac{1}{2}$,0) (blue), and (b) on the ${\bm q}=(q_1,q_2,0)$ (bottom) and ${\bm q}=(q_1,q_2,\frac{1}{2})$ (top) planes.}
\label{fig:2} 
\end{figure}

{\it Results for GdRu$_2$Si$_2$.}
Figure \ref{fig:1} shows (a) the crystal structure, (b) band structure, and (c) partial density of states (DOS) of GdRu$_2$Si$_2$. The energy eigenvalues of SDFT and the tight-binding model show good agreement in the range from $-6$ to $3$ eV. The fully polarized Gd-4$f$ orbitals are located at around $-4$ eV (majority spin) and $1$ eV (minority spin). The dominant contribution of DOS at the Fermi level comes from the Ru-4$d$ orbital, which is about 4 times as large as that of the Gd-5$d$ and Si-3$p$. In this paper, we adopt the conventional unit cell of the body-centered-tetragonal structure given in Fig.~\ref{fig:1}(a) and measure ${\bm q}$-vectors in the unit of the corresponding reciprocal lattice vectors. 

Based on the obtained tight-binding model, we calculate $J({\bm q})$ using Eq.~\eqref{eq:lich}. Figure \ref{fig:2} shows $J({\bm q})$ (a) along  high symmetry lines and (b) on the $q_z=0$ and $\frac{1}{2}$ planes. We observe the two peaks at around ${\bm q}_1^*=(0.19,0,0)$ and ${\bm q}_2^*=(0.70,0,0)$, of which the former is very close to the experimental modulation vector ${\bm Q}=(0.22,0,0)$~\cite{Seki2020}. The corresponding N\'eel temperature $T_N=45$ K is also very close to the experimental value ($46$ K). From the results, we can expect that our first-principles calculation correctly captures basic properties of the spin interactions in GdRu$_2$Si$_2$. Note that the spin texture of ${\bm q}_2^*$ also show the similar helical structure if we only focus on the in-plane modulation~\cite{supple}. Thus, the existence of nearly degenerate two peaks indicates that the out-of-plane spin correlation is not significantly strong in this material. In the following, we do not focus on the second peak at ${\bm q}_2^*$ and discuss the mechanism to stabilize the peak at ${\bm q}_1^*$ in detail.

{\it Mechanism of the peak at ${\bm q}_1^*$.} 
To identify the driving force leading to the finite-$\bm Q$ magnetic structure, we take the following steps: Since $J_{ij}$ in Eq.~\eqref{eq:lich} can be decomposed into each atomic component, we first approximate $J_{ij}$ by that of the Gd components. This procedure is justified because $\Sigma^{i0}$ of Ru-$4d$ and Si-3$p$ are much smaller than that of Gd-5$d$ and Gd-4$f$. It should be noted that the typical value of $\Sigma^{i0}$ is given by the exchange splitting energy due to the magnetic order~\cite{Nomoto2020}. Here, due to the ferromagnetic coupling to Gd-4$f$, Gd-5$d$ has a large spin splitting despite its wider spread than Ru-$4d$. In the present case, a typical energy splitting of Ru-4$d$ is about 0.04 eV while that of Gd-5$d$ is 0.75 eV. Thus, even taking into account the fact that the contribution to $J_{ij}$ is proportional to DOS at the Fermi level, we can expect that the contribution of Gd-5$d$ is about 100 times larger than that of Ru-4$d$. 

\begin{figure}[t]
\centering
\includegraphics[width=8.5cm]{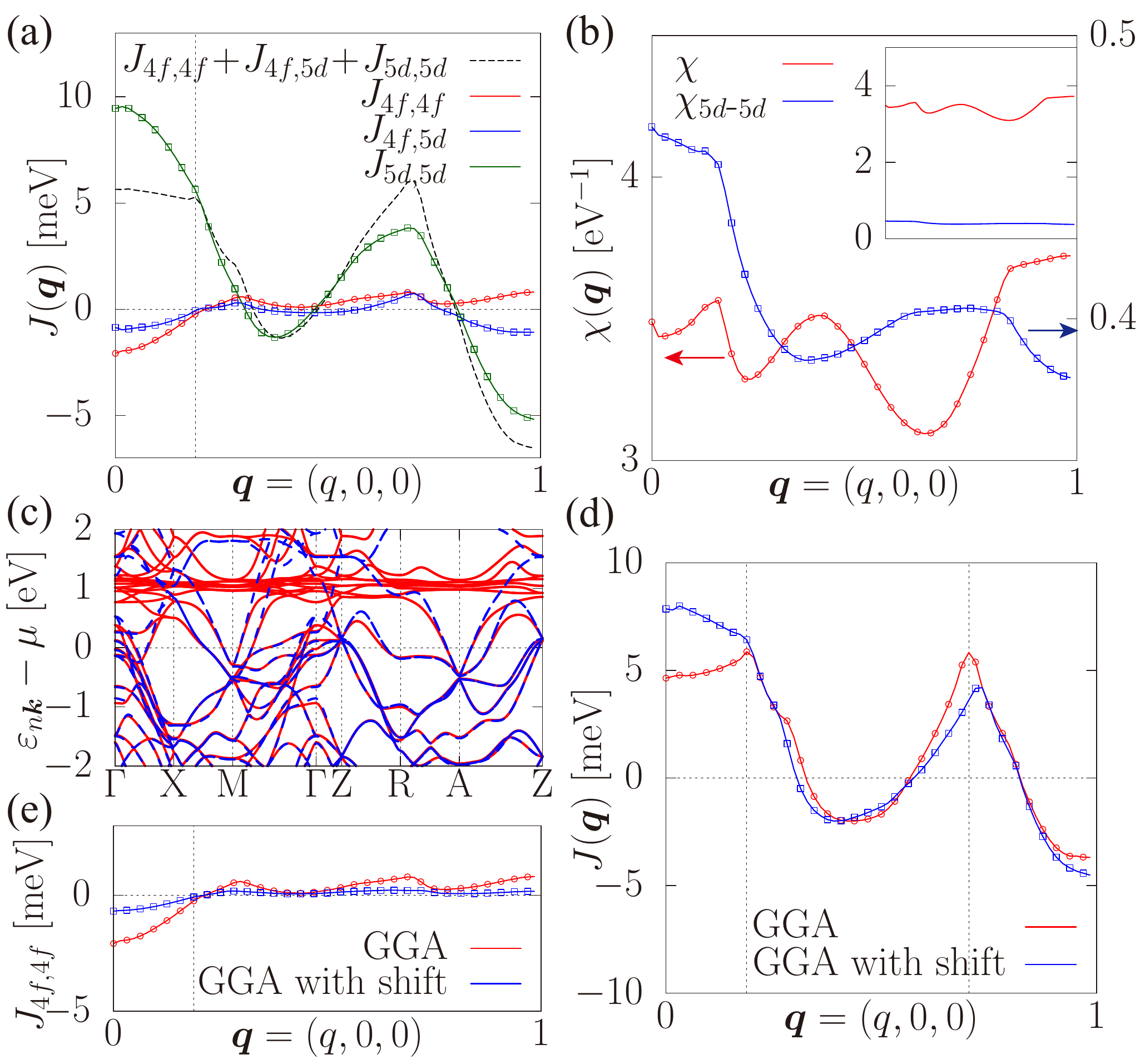}
\caption{(a) Orbital decomposed $J(\bm q)$ defined by Eqs.~\eqref{eq:dec1}-\eqref{eq:dec3}. (b) Spin susceptibility for the paramagnetic state. Red and blue lines correspond to $\chi(\bm q)$ and its Gd-$5d$ component, respectively. (c)-(e) Comparison between GGA (red) and GGA with the energy shift (blue) results: (c) The band structures of minority spin, (d) $J({\bm q})$ evaluated by Eq.~\eqref{eq:lich}, and (e) $J_{4f,4f}({\bm q})$ evaluated by Eq.~\eqref{eq:dec3}.}
\label{fig:3} 
\end{figure}

For the Gd components of $J_{ij}$, we can also decompose them into Gd 5$d$-5$d$, Gd 4$f$-4$f$ and their cross term contributions. By assuming that $\Sigma^{i0}$ is orbital independent and simply proportional to Pauli matrix $\sigma_x$, each contribution can be evaluated as follows:
\begin{align}
J^{4f,4f}_{ij}&=-\frac{\Delta_{4f}^2}{4}{\rm Tr}_{\omega_n\ell}[G_{j\uparrow i\uparrow }^{4f}G_{i\downarrow j\downarrow}^{4f}], \label{eq:dec1}\\
J^{4f,5d}_{ij}&=-\frac{\Delta_{4f}M_{5d}}{2}{\rm Tr}_{\omega_n\ell}[G_{j\uparrow i\uparrow}^{4f}G_{i\downarrow j\downarrow}^{5d}], \label{eq:dec2}\\
J^{5d,5d}_{ij}&=-\frac{\Delta_{5d}^2}{4}{\rm Tr}_{\omega_n\ell}[G_{j\uparrow i\uparrow}^{5d,\uparrow}G_{i\downarrow j\downarrow}^{5d}], \label{eq:dec3}
\end{align}
where the exchange splitting energy $\Delta_{4f}$ of Gd-$4f$ and $\Delta_{5f}$ of Gd-$5d$ are set to $0.75$ and $5.2$ eV, respectively. These three contributions to $J({\bm q})$ are shown in Fig.~\ref{fig:3}(a) along the (0,0,0)-(1,0,0) line. We can see that $J^{5d,5d}(\bm q)$ shows a peak at ${\bm q}=(0,0,0)$, which means that the interaction between 5$d$ is ferromagnetic. On the other hand, $J^{4f,4f}(\bm q)$ and $J^{4f,5d}(\bm q)$ show the opposite behavior, namely, there is a dip structure at $\bm q=(0,0,0)$. Although a slight deviation due to the ignored contributions exists, the summation over the three terms, corresponding to the black dashed line in Fig.~\ref{fig:3}(a), roughly reproduces the result of $J({\bm q})$~\cite{note02}. Thus, we can conclude that the competition between the ferromagnetic interactions in the Gd-$5d$ manifold and antiferromagnetic interactions in the Gd-$4f$ manifold is the main origin of the finite-${\bm Q}$ structure in GdRu$_2$Si$_2$. This is our main finding in the paper.  Next, we discuss physical interpretations of this behavior by considering the limiting cases of the formulas~\eqref{eq:dec1}-\eqref{eq:dec3}. 

{\it Limiting cases.}
First, let us consider $J_{ij}^{5d,5d}$ in the limit of $\Delta_{5d}, V_{4f,5d}\rightarrow 0$, where $V_{4f,5d}$ represents hybridization between Gd-$4f$ and Gd-$5d$ orbitals. In this limit, one may write $J_{ij}^{5d,5d}$ as follows:
\begin{align}
J_{ij}^{5d,5d}\sim \Delta_{5d}^2\chi^{5d,5d}_{ij}, \label{eq:dec1_}
\end{align}
where $\chi^{5d,5d}_{ij}$ denotes the Gd-$5d$ component of the spin susceptibility in the paramagnetic state~\cite{note03}. Its Fourier transform $\chi^{5d,5d}(\bm q)$ is shown by the blue line in Fig.~\ref{fig:3}(b), whose $\bm q$-dependence is essentially the same as $J^{5d,5d}(\bm q)$ in the ferromagnetic state (Fig.~\ref{fig:3}(a)). This indicates that the magnetic order of Gd-$4f$ orbitals does not affect the electronic structure of Gd-$5d$ so much. Note that although $\chi^{5d,5d}(\bm q)$ should be sensitive to the Fermi surface nesting, it is not identical to the real spin susceptibility $\chi(\bm q)$. This is because $\chi(\bm q)$ includes the contributions from all atoms on an equal footing, which is shown by the red line of Fig.~\ref{fig:3}(b). 
If we consider a fictitious Hubbard-like model for the Gd-$5d$ orbitals, we can regard $J^{5d,5d}$ as an effective (screened) interaction where $\Delta_{5d}$ corresponds to the bare $U$. 
%Since the effective interaction induces conventional SDW instability, 
Thus the ferromagnetic feature of $J^{5d,5d}$ can be understood in terms of the $\bm q$-dependence of the %three-dimensional
Lindhard function of the fictitious model.

Next, let us consider the antiferromagnetic interaction $J^{4f,4f}$. Here, we can take two different limits: One corresponds to the limit of $V_{4f,5d}\rightarrow0$ keeping $t_{4f}$ finite, where $t_{4f}$ represents the direct hopping integral between the Gd-$4f$ orbitals, and the other corresponds to the opposite limit. After taking the limit of $\Delta_{4f}\rightarrow \infty$, one obtains,
\begin{align}
J_{{\rm SR},ij}^{4f,4f}&=J_{ij}^{4f,4f}|_{V_{4f,5d}\rightarrow 0}\sim-\frac{t_{4f}^2}{\Delta_{4f}}, \label{eq:dec3_}\\
J_{{\rm RKKY},ij}^{4f,4f}&=J_{ij}^{4f,4f}|_{t_{4f}\rightarrow0}\sim J_K^2\chi^{5d,5d}_{ij},
\end{align}
which respectively represents the short-range interaction between the Gd-$4f$ electrons and the RKKY long-range interaction mediated by the Gd-$5d$ electrons, where $J_K=V_{4f,5d}^2/\Delta_{4f}$ corresponds to the Kondo coupling strength. Here, we consider only the super-exchange interaction originated from direct hopping between the Gd-$4f$ orbitals. While there may also be the super-exchange interaction mediated by the Si-$3p$ and Ru-$4d$ orbitals, it will not change the following discussion.

It is interesting to note that the RKKY interaction $J_{{\rm RKKY}}^{4f,4f}$ makes almost no effect on $J({\bm q})$, which can be seen from the following observations: Even though the mechanisms are different, $J_{{\rm RKKY}}^{4f,4f}$ and $J^{5d,5d}$ have the same $\bm q$-dependence determined by $\chi^{5d,5d}$. Thus, if $J_{{\rm RKKY}}^{4f,4f}$ has the dominant effect, $\bm q$-dependence of $J^{4f,4f}$ must be similar to that of $J^{5d,5d}$, which is inconsistent with the results shown in Fig.~\ref{fig:3}(a). The irrelevance of $J_{{\rm RKKY}}^{4f,4f}$ can be also understood by comparing the factors, $\Delta_{5d}^2$ in $J^{5d,5d}$ and $J_K^2$ in $J_{{\rm RKKY}}^{4f,4f}$. If we take the values $(V_{4f,5d},\Delta_{5d},\Delta_{4f})\sim (0.10, 0.75,5.20)$~eV from the tight-binding model, we find that the ratio $J_{{\rm RKKY}}^{4f,4f}/J^{5d,5d}$ becomes less than $10^{-5}$.  These results strongly suggest that the RKKY interaction essentially has no effect in determining the modulation vector ${\bm Q}$. Note that, in rare-earth compounds, it is widely believed that the RKKY interaction dominates the magnetic structure in the limit of $J_K\rightarrow 0$ since the Kondo effect screens the local spin moments in the opposite limit~\cite{Doniach}. This physical picture, however, is obtained based on the Kondo model, which only includes the exchange coupling between the conduction and the local spins. Our {\it ab initio} theory indicates that the following generalized Kondo model is better to describe the physics in GdRu$_2$Si$_2$:
\begin{align}
    \cH=\cH_{\rm Kondo}+U_c\sum_i n^c_{i\uparrow} n_{i\downarrow}^c+\sum_{ij}J_{ij}{\bm S}_i\cdot{\bm S}_j,
\end{align}
where the first term represents the Kondo Hamiltonian $\cH_{\rm Kondo}=\sum_{ij\sigma}t_{ij}^c c^\dagger_{i\sigma} c_{j\sigma}+J_{K}\sum_i {\bm s}_i^c\cdot {\bm S}_i$. In the limit of $J_K\rightarrow0$, the remaining two terms, the Coulomb interaction between the conduction electrons, which leads to $J^{5d,5d}$ in Eq.~\eqref{eq:dec1_}, and the short-range interaction between the local spins, which corresponds to $J^{4f,4f}_{\rm SR}$ in Eq.~\eqref{eq:dec3_}, will dominate the magnetic structure.
%Our {\it ab initio} calculation indicates that such a naive model cannot correctly capture the physics in real rare-earth compounds. Namely, in nature, the Coulomb interaction between the $5d$-electrons and Hund's coupling between the $5d$ and $4f$ electrons always exist, which can induce conventional SDW instability in the $5d$-electrons and will dominate the magnetic structure in the limit of $J_K\rightarrow0$.

\begin{figure}[t]
\centering
\includegraphics[width=8.5cm]{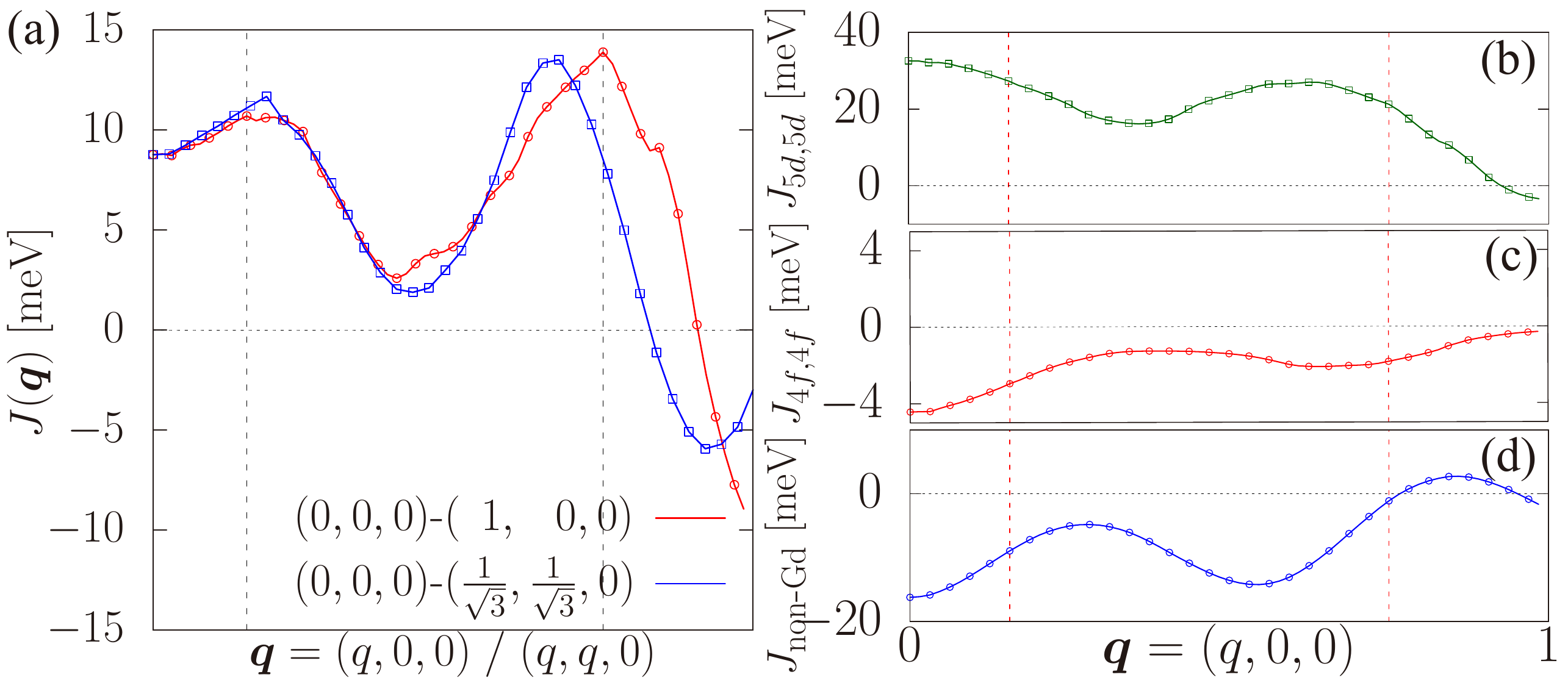}
\caption{(a) The momentum dependence of $J(\bm q)$ in Gd$_2$PdSi$_3$ along the symmetry lines (0,0,0)-(1,0,0) (red) and (0,0,0)-($\frac{1}{\sqrt{3}},\frac{1}{\sqrt{3}}$,0) (blue). (b-d) Orbital decomposed $J(\bm q)$ along the (0,0,0)-(1,0,0) line. (b) Gd-$5d$, (c) Gd-$4f$ orbital contributions, and (d) other contributions including non-magnetic Pd and Si.}
\label{fig:4} 
\end{figure}

To validate the above competition scenario, we also perform the calculation by adding $1.5$~eV energy shift to the minority spin of Gd-$4f$ levels. The results are shown in Figs.~\ref{fig:3}(c)-(e). We can see that, although the Fermi surface does not change by the shift, $J({\bm q})$ shows clearly different $\bm q$-dependence from the GGA results. Most importantly, the peak at ${\bm q}_1^*=(0.19,0,0)$ disappears and the resulting $J(\bm q)$ looks similar to $J^{5d,5d}(\bm q)$ in Fig.~\ref{fig:3}(a), which is consistent with the above scenario: Since the SDW-type instability of Gd-$5d$ is determined by the shape of the Fermi surface, it does not change by the shift of $4f$ levels. On the other hand, the short-range interactions between the Gd-$4f$ electrons should be proportional to $1/\Delta_{4f}$ in the limit of $\Delta_{4f}\rightarrow \infty$. As a result, the ${\bm q}$-dependence of $J^{4f,4f}$ becomes small, as is shown in Fig.~\ref{fig:3}(e), and $J({\bm q})$ becomes ferromagnetic like $J^{5d,5d}({\bm q})$ in GGA. 
%Although the GGA+U calculation fails to reproduce the correct finite-${\bm Q}$ structure, we do not discuss it here since the estimation of the reasonable $U$ in SDFT is a quite sensitive problem.

{\it Results for Gd$_2$PdSi$_3$.}
Finally, we show the results for Gd$_2$PdSi$_3$. Since the crystal structure of Gd$_2$PdSi$_3$ is complex due to the $z$-axis stacking of the different types of the Pd-Si layers~\cite{Tang2011}, 
%it cannot be treated in the first-principles calculations. Thus, here, 
here we employ a simplified structure with four Gd-atoms in the unit cell~\cite{supple, MaxLat}. We perform the same calculations as that for GdRu$_2$Si$_2$ and obtain $J({\bm q})$ shown in Fig.~\ref{fig:4}(a). Although we find a peak structure at ${\bm q}_1^*=(0.16, 0,0)$, which is close to the experimental ${\bm Q}=(0.14,0.0)$, this peak is sub-dominant and the corresponding $T_N=82$ K is much higher than the experimental $T_N$ (21 K)~\cite{Kurumaji2019}. This might be due to the approximation for the crystal structure. However, if we look into the decomposed $J_{ij}$, we can see the same trend as GdRu$_2$Si$_2$. Namely, $J^{5d,5d}$ shows a ferromagnetic interaction with a peak at ${\bm q}=(0,0,0)$ (Fig.~\ref{fig:4}(b)), and $J^{4f,4f}$ shows an antiferromagnetic interaction with a dip at ${\bm q}=(0,0,0)$ (Fig.~\ref{fig:4}(c)). In contrast to GdRu$_2$Si$_2$, the interactions coming from non-magnetic atoms also has a large contribution to $J(\bm q)$ and shows a dip structure similar to $J^{4f,4f}$ (Fig.~\ref{fig:4}(d)). Thus, also in Gd$_2$PdSi$_3$ case, the finite-$\bm Q$ structure comes from a competition between the SDW type ferromagnetic interaction and other antiferromagnetic interactions, but the RKKY interaction does not affect the $\bm q$-dependence of $J(\bm q)$. From these calculations, we expect that the present mechanism is ubiquitous in the Gd-based skyrmion crystals. 

{\it Conclusion.}
In this paper, we investigate the spin structure in GdRu$_2$Si$_2$ and Gd$_2$PdSi$_3$ using the {\it ab initio} local force method. We decompose the obtained exchange interactions $J_{ij}$ into each orbital component and perform the spin susceptibility calculations. Our calculations show that the finite-$\bm Q$ structures in these compounds are stabilized by the competition between the ferromagnetic interaction in the Gd-5$d$ manifold, leading to the SDW instability, and the antiferromagnetic short-range interaction in the Gd-4$f$ manifold. The Fermi surface nesting is not the origin of the finite-$\bm Q$ structure, in contrast to the %naive assumption in the 
RKKY mechanism of the skyrmion crystal formations. While we studied only the modulation vector ${\bm Q}$ of the helical spin states, the inter-orbital frustration mechanism will pave a way to future {\it ab initio} materials design of the skyrmion lattice. %the present study will pave a way to {\it ab initio} materials design of the skyrmion lattice. 

{\it Acknowledgement.}
We are grateful to S. Seki, M. Hirschberger, K. Ishizaka, T. Hanaguri, J. Otsuki, M.-T. Suzuki, and S. Hayami for many valuable discussions. This work was supported by a Grant-in-Aid for Scientific Research (No.\ 19K14654, No.\ 19H05825, No.\ 19H00650, No.\ 18K03442, and No.\ 16H06345) from Ministry of Education, Culture, Sports, Science and Technology, and  CREST (JPMJCR18T3) from the Japan Science and Technology Agency.


\begin{thebibliography}{99}
\bibitem{note02} Among the neglected contributions, the cross term between Gd and the other non-magnetic atoms has the biggest contribution to $J({\bm q})$, which slightly enhances the dip structure at ${\bm q}=(0,0,0)$.
\bibitem{note03} For the paramagnetic case, we first obtain the tight-binding model based on the DFT calculation. Then, we neglect Gd-$4f$ orbitals and calculate the susceptibility.
\end{thebibliography}
\end{document}